\documentclass[letterpaper,10pt]{article}
\usepackage{osameet2}

\usepackage{amsmath,amssymb}

\usepackage[pdftex,colorlinks=true,bookmarks=false,citecolor=blue,urlcolor=blue]{hyperref} 

\begin{document}

\title{Correlated Non-Linear Phase Noise in Multi-Subcarrier Systems}

\author{Ori Golani$^1$, Dario Pilori$^2$, Gabriella Bosco$^2$, and Mark Shtaif$^1$}
\address{$^1$ School of Electrical Engineering, Tel Aviv University, Tel Aviv, Israel 69978 \\
$^2$ DET, Politecnico di Torino, 10129 Torino, Italy.}
\email{ori.golani@post.tau.ac.il}

\begin{abstract}
We explore correlated nonlinear phase noise (NLPN) in multi-subcarrier systems. We derive an analytical model for predicting the covariance between the NLPN affecting different subcarriers, and offer a simple algorithm which uses the correlations for improved NLPN mitigation. 
\end{abstract}


\vspace{-0.5cm}
\section{Introduction}
Reduction of symbol rate, i.e. subcarrier multiplexing (SCM), is a powerful technique that allows optimizing the system with respect to fiber Non-Linear Interference Noise (NLIN) \cite{poggiolini2017recent}, and reducing certain aspects of transceiver complexity \cite{qiu2014digital}. Yet, although the overall NLIN power reduces when optimizing the number of subcarriers, a greater portion of it is due to Non-Linear Phase Noise (NLPN) \cite{pilori2018residual}.
In addition, the use of Probabilistic Shaping (PS), which makes the modulation more 'Gaussian-like' \cite{fehenberger2016probabilistic}, is expected to further increase the amount of generated NLPN \cite{Dar2014Time}.

It has been shown that the relatively long correlation times that characterize NLPN can be used in order to effectively mitigate it by means of adaptive equalization \cite{golani2016modeling}. However, SCM reduces the correlation length (in terms of number of symbols), making the use of adaptive equalization more challenging.
A remedy to this situation is to take advantage of the fact that all subcarriers are jointly processed by the same receiver, and hence correlations between them can be taken advantage of. 

In this paper we demonstrate the existence of strong correlations between the NLPN of different subcarriers in SCM systems. We derive an analytical expression for the correlation and validate it with numerical simulations. 
We report of an initial attempt to exploit these correlations by means of 
an RLS-based joint processing algorithm, which is already producing encouraging results. It is expected that customized algorithms aiming at exploiting the correlations between subcarriers will significantly improve system performance.

\vspace{-0.2cm}
\section{Estimating NLPN correlations}
\begin{figure}[t]
  \centering
  \includegraphics[trim={0cm 12cm 0cm 0cm},clip,width=16cm]{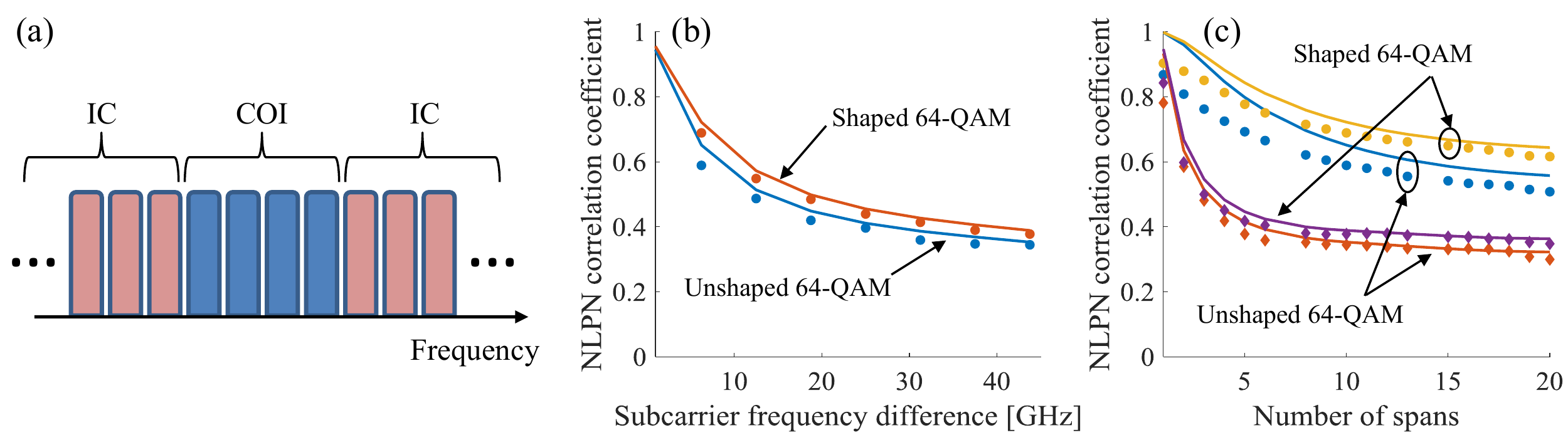}
\caption{Correlation coefficient of the NLPN affecting different subcarriers. (a) Layout of multi-subcarrier system. (b) Correlation between the lowest frequency subcarrier and all others subcarriers in the COI, for a $10\times100$km link. (c) Correlation as a function of the link length. Top two curves (circular markers) show the correlations between two adjacent subcarriers, whereas the bottom two curves (diamond markers) show the correlations between the two farthest subcarriers in the COI.}
\label{fig:correlation}
\end{figure}

The phase-noise $\phi_n^{(j)}$ affecting the $n$-th symbol of the $j$-th subcarrier in the channel of interest (COI) as a result of a single interfering channel (IC), is given by \cite{golani2016modeling}
\begin{align}
    \phi_n^{(j)}&= \sum_{k,m} X_{k,m}^{(j)} b_{n-k}^*b_{n-m}
,\end{align}
where $b_n$ are the symbols of the IC, and $X_{k,m}^{(j)}$ are the interaction coefficients (which are independent of the signal power or properties). Since the subcarriers are spectrally close to one another they are affected by the IC in a similar way and hence the coefficients $X_{k,m}^{(j)}$ are similar for different subcarriers, producing a correlation between subcarriers' NLPN. It can be shown that 
the covariance between the NLPN of the $i$-th and $j$-th subcarriers is given by
\begin{align}\label{eq:cov}
    \mathrm{Cov}(i,j)=\langle\phi_n^{(i)}\phi_n^{(j)}\rangle-\langle\phi_n^{(i)}\rangle\langle\phi_n^{(j)}\rangle = P^2\left[ S_1^{(i,j)} + (M-2)S_2^{(i,j)} \right]
,\end{align}
where $P$ is the launch power of the IC, $M$ is the normalized kurtosis of the IC's constellation \cite{dar2013properties}, and
\begin{align}
  S_1^{(i,j)}= \sum_{k,m}  X_{k,m}^{(i)} X_{k,m}^{(j)*}, \hspace{1cm} 
    S_2^{(i,j)}= \sum_{k}  X_{k,k}^{(i)} X_{k,k}^{(j)*}
.\end{align}
The values of $S_1^{(i,j)}$ and $S_2^{(i,j)}$ can be found analytically, using a similar derivation to that performed in \cite{golani2016modeling}. Once these quantities are evaluated, the covariance matrix characterizing the NLPN of all subcarriers in the WDM channel can be easily expressed, for every symbol constellation and launch power.

A split-step Fourier simulation was set up to validate the analytical derivation. The transmitter generates 21 pol-muxed WDM channels
on a 50-GHz DWDM grid. Each channel is then divided into $N_\textup{s}$ subcarriers, each with symbol rate $32/N_\textup{s}$
GBaud and separated by $50/N_\textup{s}$ GHz. As a COI, we picked the central channel in the WDM spectrum,
as shown in Fig. \ref{fig:correlation}a. Transmitted modulation formats are (unshaped) 64-QAM, and PS (shaped) 64-QAM, 
shaped with a Maxwell-Boltzmann distribution to give a constellation entropy of $5$ bit/symb in each polarization.
The kurtosis of the unshaped and shaped constellations was $M=1.38$ and $M=1.89$, respectively. 
The link consisted of identical $100$-km spans of SMF ($\alpha=0.2~\mathrm{dB}/\mathrm{km}$, 
$\beta_2=-21.27~\mathrm{ps}^2/\mathrm{km}$, $\gamma=1.3~(\mathrm{W}\cdot\mathrm{km})^{-1}$, without PMD), separated by amplifiers that compensate for the span loss. Chromatic dispersion compensation and polarization demultiplexing where performed digitally.
The NLPN covariance was extracted using the method reported in \cite{golani2018experimental}, where the NLPN of the $j$-th subcarrier is evaluated as 
\vspace{-0.2cm}
\begin{align}
    \Tilde{\phi}_n^{(j)}=\mathrm{Im}\left[\frac{s_n^{(j)} - a_n^{(j)}}{a_n^{(j)}}\right]
,\end{align}\vspace{-0.2cm}

\noindent where $s_n^{(j)}$ and $a_n^{(j)}$ are the received signal and transmitted data symbol, respectively, and $\mathrm{Im}[\cdot]$ means taking the imaginary part of the argument. The covariance between $\phi_n^{(i)}$ and $\phi_n^{(j)}$ is found by directly evaluating the empirical covariance of $\Tilde{\phi}_n^{(i)}$ and $\Tilde{\phi}_n^{(j)}$.

Results, comparing simulations (markers) with analytical predictions (solid lines), 
are shown in Fig. \ref{fig:correlation}b and \ref{fig:correlation}c in the exemplary case of 8 subcarriers. Fig. \ref{fig:correlation}b
shows the correlation between the lowest-frequency subcarrier of the COI and the other subcarriers, after 10 spans.
It can be seen that the correlation decreases with frequency, but is always  above $0.3$. Shaping mildly increases NLPN correlation with respect to unshaped
64-QAM. Fig. \ref{fig:correlation}c shows the correlation between the first and second subcarriers (top two curves) and the first and last subcarriers (two bottom curves) as a function of
the link length. In both cases, the correlation reduces with the number of spans, where the reduction is faster in the case of the larger frequency separation. Although the analytic expression slightly overestimates the actual correlation at short distances, the overall agreement between theory and simulations is self evident. 

\vspace{-0.2cm}
\section{Joint mitigation of NLPN}
\begin{figure}[t]
  \centering
  \includegraphics[trim={0cm 12cm 0cm 0cm},clip,width=16cm]{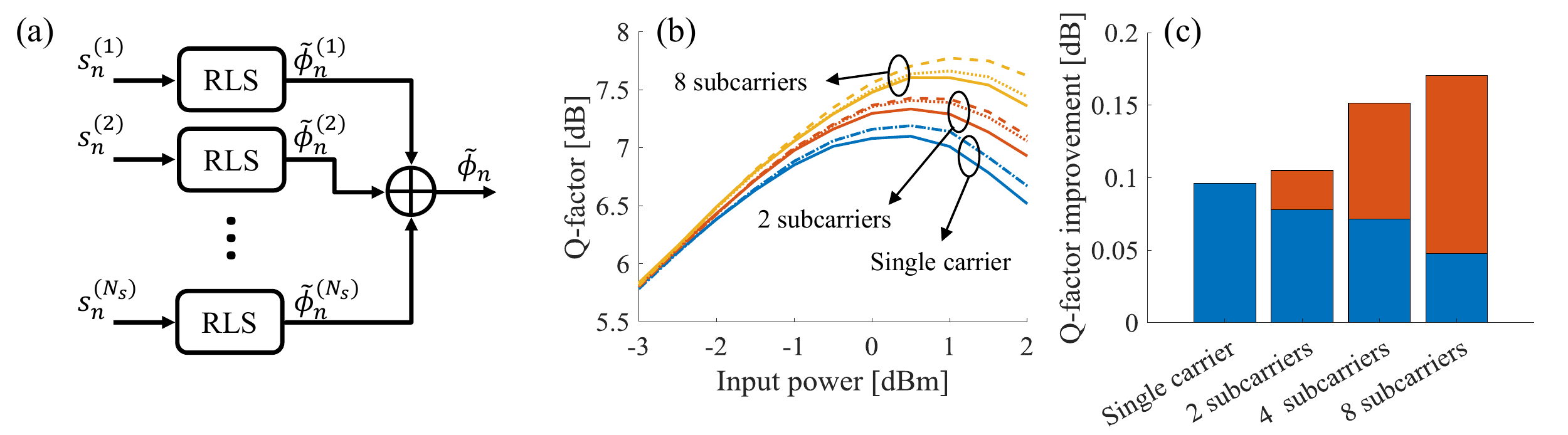}
\caption{Benefit of joint-equalization of multiple subcarriers. (a) Schematic description of the joint phase estimation algorithm. (b) Q-factor vs. launch power for a $5\times100$km link, using different number of subcarriers and equalizer types. Solid curves show the Q-factor without any equalization, dotted lines show results with individual equalization for each subcarrier, and dashed lines represent the case of joint phase estimation. (c) Peak Q-factor gain with different equalizers. Blue bars show the gain from individual equalization, whereas the red bars show the added gain of joint processing.}
\label{fig:mitigation}
\end{figure}

The main importance of the correlation between the NLPN affecting different subcarriers is the ability to employ joint phase cancellation algorithms to improve performance. As an initial demonstration of this principle, we used a simple joint phase estimation algorithm, described schematically in Fig. \ref{fig:mitigation}a. Each of the subcarriers is individually detected, and a standard decision-directed RLS equalizer is used to estimate the NLPN. The outputs of the $N_s$ equalizers are then averaged, providing a single estimation of the NLPN, which is less noisy than those of the individual equalizers. The averaged phase estimation is then removed from each subcarrier's signal, providing a cleaner estimate of the transmitted data. 

The joint equalization procedure was applied to an exemplary scenario of a $5\times100$km link, carrying unshaped 64-QAM constellation and using 11 WDM channels. The EDFAs had a noise figure of 5dB, and all other parameters are as described in the previous section. The RLS equalizers' forgetting factor was optimized independently for each configuration, so as to provide optimal performance. Figure \ref{fig:mitigation}b shows the Q-factor of the equalized signal, as a function of the launch power and the number of subcarriers. The differences between the un-equalized cases (solid curves) originate from the reduction in NLIN power at lower symbol rates. Figure \ref{fig:mitigation}c shows the peak Q-factor gain (i.e. the difference between the peak of the unequalized case to that of the equalized cases), using both individual equalization (blue bars) and joint equalization (red bars). It is evident that the benefit of individual equalization decreases as the number of subcarriers increases. This is because the temporal correlation length of the NLPN decreases, which makes the equalization more difficult. In contrast, the benefit from joint processing increases with the number of subcarriers. 

\vspace{-0.2cm}
\section{Conclusion}
We have demonstrated the existence of correlations between the NLPN of different subcarriers in a  multi-subcarrier system. The correlations were found to be fairly strong and an accurate analytic approach to finding them has been presented. We have shown a simple equalization scheme for benefiting from these correlations, and argue that much larger benefits should be achievable in equalizers specifically designed to target this feature.


\end{document}